\documentclass[twocolumn,showpacs,superscriptaddress,preprintnumbers,amsmath,amssymb,prb,floatfix]{revtex4}

\usepackage{graphicx}
\usepackage{dcolumn}
\usepackage{bm}
\usepackage{amsmath}
\usepackage{amssymb}
\usepackage{booktabs}

\begin{document}

\preprint{APS/123-QED}

\title{First principles quasiparticle damping rates in bulk lead}

\author{X. Zubizarreta}
   \affiliation{Donostia International Physics Center (DIPC), Paseo de Manuel Lardizabal 4, 20018
   San Sebasti\'an/Donostia, Basque Country, Spain}
   \affiliation{Departamento de F\'{\i}sica de Materiales, Facultad de Ciencias Qu\'{\i}micas,
   Universidad del Pa\'{\i}s Vasco/Euskal Herriko Unibertsitatea, Apdo. 1072, 20080 San Sebasti\'an/Donostia,
   Basque Country, Spain}
\author{V. M. Silkin}
   \affiliation{Donostia International Physics Center (DIPC), Paseo de Manuel Lardizabal 4, 20018
   San Sebasti\'an/Donostia, Basque Country, Spain}
   \affiliation{Departamento de F\'{\i}sica de Materiales, Facultad de Ciencias Qu\'{\i}micas,
   Universidad del Pa\'{\i}s Vasco/Euskal Herriko Unibertsitatea, Apdo. 1072, 20080 San Sebasti\'an/Donostia, Basque Country,
Spain}
   \affiliation{IKERBASQUE, Basque Foundation for Science, 48011 Bilbao, Spain}
\author{E. V. Chulkov}
   \affiliation{Donostia International Physics Center (DIPC), Paseo de Manuel Lardizabal 4, 20018
   San Sebasti\'an/Donostia, Basque Country, Spain}
   \affiliation{Departamento de F\'{\i}sica de Materiales, Facultad de Ciencias Qu\'{\i}micas,
   Universidad del Pa\'{\i}s Vasco/Euskal Herriko Unibertsitatea, Apdo. 1072, 20080 San Sebasti\'an/Donostia,
   Basque Country, Spain}
   \affiliation{Centro de F\'{\i}sica de Materiales CFM - Materials Physics Center MPC, Centro Mixto CSIC-UPV/EHU,
   Paseo de Manuel Lardizabal 5, 20018 San Sebasti\'an/Donostia, Basque Country, Spain}
\date{\today}

\begin{abstract}
First principles calculations of the damping rates (inverse inelastic lifetimes), $\tau^{-1}$, of low energy
quasiparticles in bulk Pb are presented. Damping rates are obtained both for excited electrons and holes with
energies up to 8 eV on a set of \textbf{k} vectors throughout the Brillouin zone (BZ). Strong localization
effects in the calculated $\tau^{-1}$ are found. Averaged over the BZ inelastic lifetimes versus quasiparticle
energy are reported as well. In addition, the effect of the spin-orbit induced splitting in the band structure on
the calculated lifetimes in Pb is investigated.
\end{abstract}


\maketitle

\section{INTRODUCTION}

In metals electron-electron inelastic scattering processes give rise to the main contribution to the damping rate
of excited electrons (and holes) with energies $\gtrsim$0.5 eV above (below) the Fermi level. For a long time the
basic knowledge for such kind of processes was based on theories developed for a free electron gas
(FEG).\cite{qufepr58,qupr62,no62} Recently, when the calculations from the first principles became to be
computationally feasible, this field has experienced profound modifications. Thus the first principles
calculations \cite{capiprl99,sckeprb99,kescprb00,caruprl00,capiprb00,gebeprb01,ladeprb03,pizhchemphys04,lahoprb04,zhchpu09,nezhpss07} have shown that the inelastic
lifetime of excited electrons indeed is a result of balance between localization, screening and band structure
details, even in metals whose electronic structure frequently considered as being a free-electron-like
one.\cite{casiprb00,neskprb08} In the previous theoretical work a non-free-electron like behavior of damping
rates in Be,\cite{casiprb00,sichprb03} the role of the screening of the $d$-electrons in inelastic lifetimes on
Cu,\cite{casiprb00,kescprb00} transient exciton\cite{ekscapa00,scijmp03} and full inclusion of
exchange-correlation (XC) effects\cite{gupiprb04} as well as renormalization effects on quasiparticle (excited
electron or hole) lifetimes in noble and transition paramagnetic\cite{zharprb01,bascprb02,scpss07} and
ferromagnetic metals\cite{zhchprl04} and compounds\cite{appphya,nechpss09,nechepjb10} were studied by means of the GW
approximation.\cite{helussp69} Using the T-matrix theory \cite{fewa71,ma90} the consequences of spin-flip
processes on damping rates in magnetic materials were found to be significant for the spin-minority
states.\cite{zhchprl04,zhanprb04,zhchprb05} In Ref. \onlinecite{zharprb02} good agreement was found between the
inelastic lifetimes for excited electrons and holes in several metals evaluated within the GW approximation and
the semiempirical scattering-theory approach. Role of inclusion of the accurate quasiparticle band structure of
Cu and Ag in the quasiparticle lifetimes was investigated.\cite{madeprb02,yimaprb10}

At the same time, in heavy elements an additional ingredient like spin-orbit (SO) interaction starts to be
important in the description of the electronic structure. A well-known example is corresponding modifications in
band structures in Bi and Pb. However, to the best of our knowledge, up to now the effect of the SO interaction
on quasiparticle lifetimes in real materials taking into account its band structure evaluated from first
principles was not investigated. In the present paper, for the first time, inelastic lifetimes of excited
electrons and holes are studied in bulk lead by means of first-principle calculations, analyzing in detail the
band structure as well as SO coupling effects. Recently thin films of Pb grown on different substrates have
obtained great deal of attention. The questions regarding growth, transport, magnetic, superconducting properties
of these systems have been considered. For instance, the confinement effects on the superconducting transition
temperature\cite{guzhs04,eoqiprl06,qikis09,brhoprl09,zhchnp10} and quasiparticle decay
rates\cite{hobrprb09,kirenp10} were studied. Recent lifetime measurements by two-photon
spectroscopy\cite{kirenp10} for few monolayers Pb deposited on silicon substrate point out to the conservation of
the bulk-like behavior of decay processes in Pb even in the very thin slabs. Therefore detailed first principles
investigation of the inelastic decay rates of quasiparticles in bulk lead which can serve as a reference for
existing and future experiments seems useful and timely.

In the present study the Kohn-Sham equations of density-functional theory \cite{hokopr64,koshpr65} are solved
self-consistently within norm-conserving pseudopotential scheme. Subsequently, linear response theory is used to
calculate the momentum- and energy-dependent density response functions over the Brillouin zone (BZ) from which
the imaginary part of the quasiparticle self-energy is computed invoking the GW approximation of many-body
theory. Comparison of the obtained results with that for the FEG model shed light on the role of the screening in
this metal. In order to study the effect of the SO coupling on the quasiparticle lifetime through its influence
on the band structure the latter was evaluated both including and excluding the SO term in the Hamiltonian.
The resent work demonstrates that the inclusion of the SO interaction into the band structure calculation
produces noticeable effect in inelastic lifetimes for quasiparticles at very low energies. Nevertheless, the
general trends in the quasiparticle inelastic lifetime as a function of its energy are unaffected by the
inclusion of the SO splitting. Additionally, comparison of the calculated inelastic lifetimes for the \textit{p}
and \textit{d} electrons with the same energy reveals strong localization effects.

The paper is organized as follows. In Sec. II, the explicit expressions for the quasiparticle decay rate in
periodic crystals within the GW approximation are presented. In Sec. III the effect of the SO interaction on the
electronic band structure of bulk Pb is discussed. The calculated results for the damping rates are analyzed in
Sec. IV along together with a link to the linewidth of the quantum well states (QWS) in Pb(111) thin films.
Finally, conclusions are given in Sec. V. Atomic units are used throughout, i.e., $e^2=\hslash=m_e=1$, unless
otherwise is stated.

\section{CALCULATION DETAILS}

In the framework of many-body theory,\cite{fewa71,ma90} for an inhomogeneous electron system, the damping rate
$\tau_i^{-1}$ of an excited electron (or hole) in the state $\phi_i(\textbf{r})$ with energy $\varepsilon_i$ is
obtained from the knowledge of the imaginary part of the quasiparticle self-energy,
$\Sigma(\textbf{r},\textbf{r}';\varepsilon_i)$, as
\begin{equation}
\label{tau_init}
\tau_i^{-1} = -2 \int {\rm d}\textbf{r} \int {\rm d}\textbf{r}' \phi_i^*(\textbf{r}) {\rm Im}
\Sigma(\textbf{r},\textbf{r}';\varepsilon_i)\phi_i(\textbf{r}').
\end{equation}

In order to calculate $\tau^{-1}$ via Eq. (\ref{tau_init}), the nonlocal energy-dependent operator
$\Sigma(\textbf{r},\textbf{r}';\varepsilon_i)$ is evaluated by means of the GW approximation,\cite{helussp69} in
which one considers only the first order term in a series expansion of the self-energy in terms of the screened
Coulomb interaction:
\begin{equation}
\Sigma(\textbf{r},\textbf{r}';\varepsilon_i) = \frac{\rm i}{2\pi}\int
G(\textbf{r},\textbf{r}';\varepsilon_i-\varepsilon)W(\textbf{r},\textbf{r}';\varepsilon){\rm d}\varepsilon,
\end{equation}
where $G(\textbf{r},\textbf{r}';\varepsilon_i-\varepsilon)$ stands for the one-particle Green function and
$W(\textbf{r},\textbf{r}';\varepsilon)$ is the dynamically screened Coulomb interaction. Replacing the Green
function by its noninteracting counterpart, the imaginary part of the self-energy can be expressed in the
following way:
\begin{equation}
\label{sigma_rr}
{\rm Im}\Sigma(\textbf{r},\textbf{r}';\varepsilon_i) = \sum_f \phi_f^*(\textbf{r}') {\rm Im}
W(\textbf{r},\textbf{r}';\omega)\phi_f(\textbf{r}),
\end{equation}
where $\omega=|\varepsilon_i-\varepsilon_f|$ represents the energy transfer and the sum is extended over a
complete set of final states $\phi_f(\textbf{r})$ with energy $\varepsilon_f$ (with constrain $E_F\leqslant
\varepsilon_f\leqslant \varepsilon_i$ for electrons and $\varepsilon_i\leqslant \varepsilon_f\leqslant E_F$ for
holes), being $E_F$ the Fermi energy. The screened Coulomb interaction is written as
\begin{multline}
\label{w_rr}
W(\boldsymbol{r},\boldsymbol{r}';\omega) = v(\boldsymbol{r}- \boldsymbol{r}') + \int {\rm
d}\boldsymbol{r}_1 \int
{\rm d}\boldsymbol{r}_2v(\boldsymbol{r}- \boldsymbol{r}_1)\\
\times\chi(\boldsymbol{r}_1,\boldsymbol{r}_2;\omega)v(\boldsymbol{r}_2- \boldsymbol{r}'),
\end{multline}
where $v(\textbf{r}- \textbf{r}')$ is the bare Coulomb interaction, and $\chi(\textbf{r},\textbf{r}';\omega)$ is
the density response function for interacting electrons.

In the framework of time-dependent density-functional theory \cite{rugrprl84,pegoprl96} and within the
random-phase approximation (RPA), the density response function $\chi$ satisfies the integral equation
\begin{multline}
\chi(\boldsymbol{r},\boldsymbol{r}';\omega) = \chi^{\rm o}(\boldsymbol{r},\boldsymbol{r}';\omega) + \int {\rm
d}\boldsymbol{r}_1 \int {\rm d}\boldsymbol{r_2}\chi^{\rm o}(\boldsymbol{r},\boldsymbol{r}_1;\omega)\\ \times
v(\boldsymbol{r_1}- \boldsymbol{r}_2)\chi(\boldsymbol{r}_2,\boldsymbol{r}';\omega),
\end{multline}
where $\chi^{\rm o}(\textbf{r},\textbf{r}';\omega)$ is the density response function of the noninteracting
electrons. Choosing the RPA one neglects the short-range XC effects in the evaluation of
$\chi(\textbf{r},\textbf{r}';\omega)$. At T=0 K, $\chi^{\rm o}(\textbf{r},\textbf{r}';\omega)$ reads
\begin{multline}
\label{xi_0} \chi^{\rm o}(\boldsymbol{r},\boldsymbol{r}';\omega) = 2 \sum_{j,j'}\frac{\theta(E_F-\varepsilon_j) -
\theta(E_F-\varepsilon_{j'})}{\varepsilon_j-\varepsilon_{j'}+(\omega+{\rm i}\eta)}\\
\times\phi_j(\boldsymbol{r})\phi_{j'}^*(\boldsymbol{r})\phi_{j'}(\boldsymbol{r}')\phi_j^*(\boldsymbol{r}').
\end{multline}

For tree-dimensional periodic crystals the following Fourier expansion for the screened interaction of Eq.
(\ref{w_rr}) can be introduced:
\begin{multline}
\label{w_gg}
W(\textbf{r},\textbf{r}';\omega) = \frac{1}{\Omega} \sum_{\textbf{q}}^{BZ}\sum_{\textbf{G},
\textbf{G}'}e^{{\rm i}(\textbf{q}+\textbf{G})\cdot\textbf{r}}e^{-{\rm i}(\textbf{q}+\textbf{G}')\cdot\textbf{r}'}\\
\times v_{\textbf{G}}(\textbf{q})\epsilon^{-1}_{\textbf{G},\textbf{G}'}(\textbf{q},\omega),
\end{multline}
where the first sum is extended over the first BZ, $\textbf{G}$ and $\textbf{G}'$ are reciprocal lattice vectors,
$\Omega$ is the normalization volume, $v_{\textbf{G}}(\textbf{q})$ represents the Fourier coefficients of the
bare Coulomb interaction, and $\epsilon^{-1}_{\textbf{G},\textbf{G}'}(\textbf{q},\omega)$ are the Fourier
coefficients of the inverse dielectric function,
\begin{equation}
\label{eps-1}
\epsilon^{-1}_{\textbf{G},\textbf{G}'}(\textbf{q},\omega) = \delta_{\textbf{G},\textbf{G}'} +
\chi_{\textbf{G},\textbf{G}'}(\textbf{q},\omega)v_{\textbf{G}'}(\textbf{q}).
\end{equation}

Introducing Eq. (\ref{w_gg}) into Eq. (\ref{sigma_rr}), the following expression for the damping rate of a
quasiparticle in the initial state $\phi_{n_i\textbf{k}}(\textbf{r})$ with energy $\varepsilon_{n_i\textbf{k}}$
is obtained:
\begin{multline}
\label{tau_final} \tau_{n_i{\bf k}}^{-1} = \frac{1}{\pi^2}\sum_{n_f} \int_{BZ}{\rm
d}\textbf{q}\sum_{\textbf{G},\textbf{G}'}
\frac{B_{n_in_f}^*({\bf k},{\bf q},{\bf G})B_{n_in_f}({\bf k},{\bf q},{\bf G}')}{|\textbf{q}+\textbf{G}|^2}\\
\times \rm{Im}[-\epsilon^{-1}_{\textbf{G},\textbf{G}'}(\textbf{q},\omega)],
\end{multline}
where $\omega=\varepsilon_{n_i\textbf{k}}-\varepsilon_{n_f\textbf{k}-\textbf{q}}$, and
\begin{equation}
\label{b}
 B_{n_in_f}({\bf k},{\bf q},{\bf G}) = \int d\textbf{r}\phi_{n_i\textbf{k}}^*(\textbf{r})e^{{\rm i}
(\textbf{q}+\textbf{G})\cdot\textbf{r}} \phi_{n_f\textbf{k}-\textbf{q}}(\textbf{r}).
\end{equation}

Note that all matrix elements of inverse dielectric function
$\epsilon^{-1}_{\textbf{G},\textbf{G}'}(\textbf{q},\omega)$ enter Eq. (\ref{tau_final}) because of the coupling
between wave vectors $\textbf{q}+\textbf{G}$ and $\textbf{q}+\textbf{G}'$ with $\textbf{G}\neq\textbf{G}'$ as a
consequence of the electron density variation in solids. These coupling terms represent so-called crystalline
local-field effects.\cite{adpr62}

For evaluation of the density response function $\chi^{\rm o}(\textbf{r},\textbf{r}';\omega)$ and the matrices
$B_{n_in_f}({\bf k},{\bf q},{\bf G})$ the eigenfunctions of a Kohn-Sham system without SO interaction are
expanded in a plane-wave basis,
\begin{equation}
\label{psi}
\phi_{n\textbf{k}}(\textbf{r}) =
\frac{1}{\sqrt{\Omega}}\sum_{\textbf{G}}u_{n\textbf{k}}(\textbf{G})e^{{\rm
i}(\textbf{q}+\textbf{G})\cdot\textbf{r}}.
\end{equation}
When the SO interaction is included into the Kohn-Sham Hamiltonian, one has to work with spinors as
eigenfunctions. In this case,
\begin{equation}
\label{psi_spinor}
\Phi_{n\textbf{k}}(\textbf{r}) =
\frac{1}{\sqrt{\Omega}}\sum_{\sigma}\sum_{\textbf{G}}u_{n\sigma\textbf{k}}(\textbf{G}) e^{{\rm
i}(\textbf{q}+\textbf{G})\cdot\textbf{r}}\xi_{\sigma},
\end{equation}
where
$\xi_{\uparrow}=\frac{1}{\sqrt{2}}\binom{1}{0}$
 for spin up and
$\xi_{\downarrow}=\frac{1}{\sqrt{2}}\binom{0}{1}$ for spin down.

In this work the electron-ion interaction was represented by norm-conserving nonlocal
pseudopotential,\cite{trmaprb91} and the local-density approximation was chosen for the exchange and correlation
potentials in the Ceperley-Alder form\cite{cealprl80} using the Perdew-Zunger parametrization.\cite{pezuprb81}
Well-converged results for the band structure have been obtained with a energy cutoff of 14 Ry which corresponds
to the inclusion of $\sim$180 plane waves in the expansion of the Bloch states.

The calculations of the inverse dielectric matrices through Eq. (\ref{eps-1}) were carried out using for
evaluation of $\chi^{\rm o}$ in Eq. (\ref{xi_0}) a $\textbf{k}$ sampling over $\approx32000$ vectors in the
irreducible part of BZ (IBZ) and including 25 energy bands. The broadening parameter $\delta$, employed in the
evaluation of $\chi^{\rm o}$ as explained e.g. in Refs. \onlinecite{arguprb94} and \onlinecite{sichprb09}, was
set to 10 meV. In the expansion of dielectric matrices 40 plane waves have been considered. The sums over
reciprocal vectors $\textbf{G}$ and $\textbf{G}'$ in Eq. (\ref{tau_final}) have been extended over 40 vectors as
well.

\section{SPIN-ORBIT EFFECTS IN PB BAND STRUCTURE}

\begin{figure}
\includegraphics[width=0.47\textwidth]{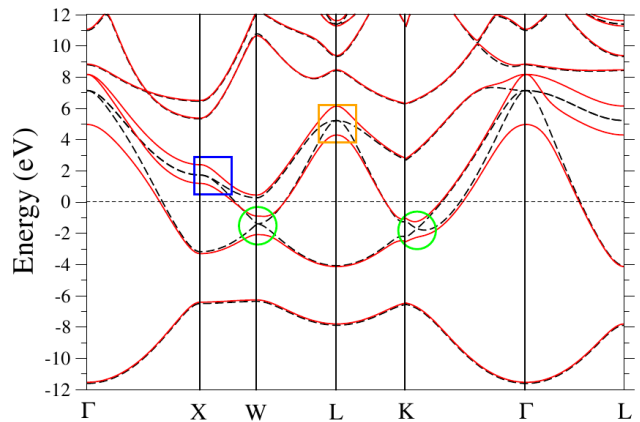}
\caption{(Color online) Calculated band structure of bulk lead, with (solid lines) and without (dashed
lines) inclusion of the spin-orbit coupling. Squares and circles mark regions where the SO-induced band splitting
is reflected in the density of states shown in Fig. \ref{dos}. The horizontal dotted line represents the Fermi
level.} \label{Pb_band_structure}
\end{figure}

\begin{table}[b]
\caption{Energies $\varepsilon_i$ of $p$-like electronic states at some high-symmetry points in the Brillouin
zone obtained in scalar-relativistic (sc) {\it ab initio} calculations and with SO interaction included computations (so). All energies
are in eV with respect to the Fermi level.}
\begin{tabular}{rrrrrrrrrrrrr}
\hline\hline\noalign{\smallskip}
 &\multicolumn{2}{c}{$\Gamma$}&\multicolumn{2}{c}{$X$}&\multicolumn{2}{c}{$W$}&\multicolumn{2}{c}
 {$L$}&\multicolumn{2}{c}{$K$}\\
\cline{2-3}\cline{4-5}\cline{6-7}\cline{8-9}\cline{10-11}\noalign{\smallskip} & sc & so
& sc & so & sc & so &  sc &  so & sc &  so \\
\hline\noalign{\smallskip}
$\varepsilon_1$&7.14&4.97&-3.18&-3.30&-1.40&-2.09&-4.07&-4.13&-2.25&-2.52\\
$\varepsilon_2$&7.14&8.18&1.73&1.20&-1.40&-0.89&5.23&4.30&-1.19&-1.04\\
$\varepsilon_3$&7.14&8.18&1.73&2.40&0.28&0.45&5.23&6.15&2.70&2.79\\
\hline\hline
\end{tabular}\label{table_1}
\end{table}

Bulk Pb lattice has a face-centered-cubic (fcc) crystal structure. Fig. \ref{Pb_band_structure} presents the
calculated band structure for lead along the high-symmetry directions of the BZ obtained with the use of the
experimental lattice parameter $a_c=4.95$ $\mathring{\rm A}$. In this figure one set of data (red solid lines)
corresponds to the calculation with the inclusion of the SO interaction in the Kohn-Sham Hamiltonian whereas
black dashed lines present results with the exclusion of SO interaction. As the fcc lattice has inversion symmetry,
due to the Kramers degeneracy \cite{ti71} each energy band is at least double degenerate in spin in both cases.
The calculated band structure is in good agreement with other theoretical results\cite{vetoprb08} and with the
experimental data\cite{jepoprb90} when the SO term is taken into account. As can be seen in Fig.
\ref{Pb_band_structure}, the inclusion of the SO interaction affects only three $p$ energy bands (the only ones
that cross the Fermi level) mainly around the high-symmetry points. In Table \ref{table_1}, a comparison of the
energies of the $p$ bands at the high-symmetry points with the inclusion and exclusion of the SO coupling is
presented. One can see that the SO interaction produces two main effects in the Pb band structure:

\begin{figure}[t]
\includegraphics[width=0.47\textwidth]{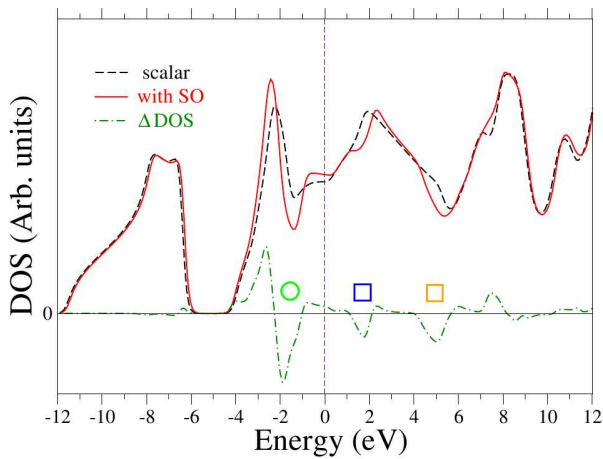}
\caption{(Color online) Total density of states (DOS) obtained in the scalar-relativistic calculation (DOS$_{\rm
SC}$, dashed line) and the calculation with inclusion of the spin-orbital (SO) term (DOS$_{\rm SO}$, solid line). The main variations in the DOS upon the SO inclusion reflected in differential $\Delta$DOS=DOS$_{\rm
SO}$-DOS$_{\rm SC}$ (dashed-dotted line) are related to different band structure splitting marked by the
same symbols as in Fig. \ref{Pb_band_structure}. DOS is in arbitrary units and energy is with respect to the Fermi
level.} \label{dos}
\end{figure}

a) At the scalar-relativistic level, at some points and
some high-symmetry lines of the BZ, the number of states in the the energy bands can be more than two (however
always it must be an even number because of the Kramers degeneracy). The SO coupling lifts some of these
``extra'' symmetry degeneracies. The largest SO-splittings $\Delta\varepsilon^{\rm SO}$ are observed for $p$ electron bands at the $\Gamma$, $X$, $L$, $W$ points and close to $K$, as well asalong the respective symmetry directions.
At the high-symmetry points we obtain: $\Delta\varepsilon^{\rm SO}=$ 3.21 eV at the
$\Gamma$ point, 1.20 eV at $X$ and $W$, and 1.85 eV at $L$ (see Table \ref{table_1}).

b) The SO interaction leads to the avoiding band-crossing effect, as the ones between $p$-like states marked by circles in Fig. \ref{Pb_band_structure}.

In Fig. \ref{dos}  the total density of states (DOS) is plotted as a function of energy. The most important
effect of inclusion of the SO interaction is the weight loss around -1.4 eV and the increase of
the peak at -2.4 eV, whose position is also slightly shifted to higher binding energies. These variations reflect
disappearance of the band-crossing points mentioned above. The other effect - breaking of the band degeneracies - has no important effect on the DOS, because
it does not flatten significantly the band dispersion. Nevertheless, the SO-induced splitting around $X$ and $L$
points affects the DOS at $\sim$1.6 eV and $\sim$5 eV, respectively
(dips in $\Delta$DOS in Fig. \ref{dos} marked by squares).

In the band structure calculations the Hamiltonian including the SO term was solved fully self consistently.
However, in the lifetime calculations the spinor representation, Eq. (\ref{psi_spinor}), for the relativistic wavefunctions is not used. First, including the full spinor structure of the relativistic wavefunctions dramatically increases the computation time. Secondly, it is well known that in simple systems with strong spin-orbit coupling, the SO-splittings of the one-electron energies are important. Nevertheless the total electronic density and furthermore the partial contribution of each one-electron state to the total density show negligible variations upon switching the SO interaction (see, e.g., Ref. \onlinecite{Bi_SO}). Hence, we suggest that one could mostly retain the SO effects on $\tau_{n_i{\bf k}}^{-1}$ just through the inclusion in Eqs. (\ref{xi_0}) and (\ref{tau_final}) of the one-electron energies calculated including the SO term in the Hamiltonian, while using the scalar-relativistic wavefunctions Eq. (\ref{psi}) in Eqs. (\ref{xi_0}) and (\ref{b}). As a result, in the SO case the energy-loss function
Im$[-\epsilon^{-1}_{\textbf{G},\textbf{G'}}(\textbf{q},\omega)]$ entering Eq. (\ref{tau_final}) for
$\tau_{n_i{\bf k}}^{-1}$ differs from that used in scalar-relativistic calculations. In a future work, lifetime calculations including the spinor representation of the relativistic wavefunctions will show if this is a meaningful approximation for including the SO effects on the calculated $\tau_{n_i{\bf k}}^{-1}$.

\begin{figure*}[t]
\includegraphics[width=1.0\textwidth]{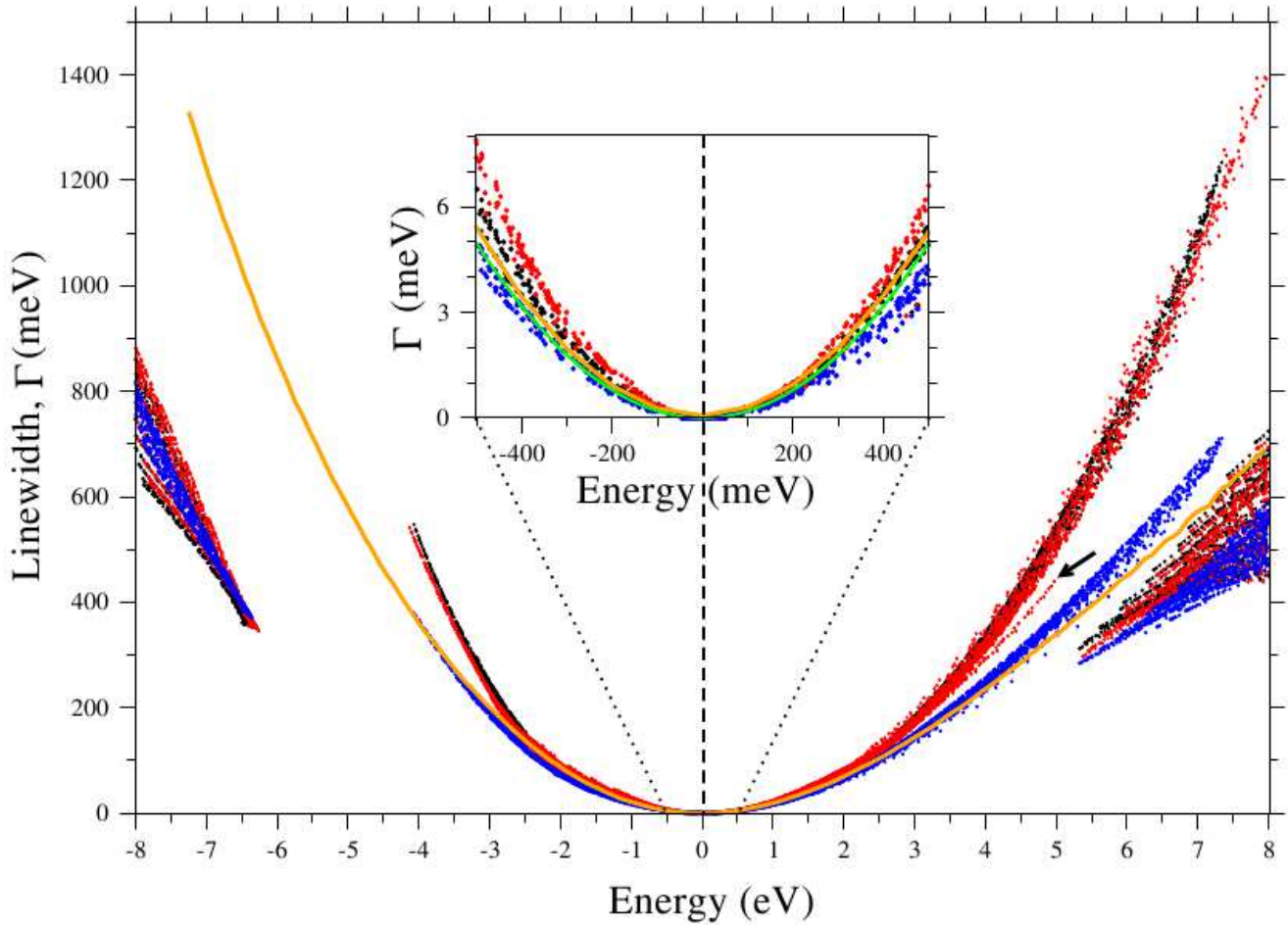}
\caption{(Color) Calculated damping rate for hot electrons and holes in all the bands and \textbf{k} points set
over the irreducible part of the Brillouin zone included in the calculations. Red dots represent the values
calculated including the spin-orbit (SO) coupling in the band structure, the black lines represent the
scalar-relativistic results and the blue lines are the results obtained using the Lindhard dielectric function
together with the scalar-relativistic band structure. The orange line shows linewidth obtained from a free
electron gas model with the Pb valence charge density parameter $r_s=2.298$ a.u. The green curve in the inset
represents results for linewidth calculated according to the Quinn-Ferrel expression according to Eq.
(\ref{gamma_QF}) for the same $r_s$. The black arrow points to the linewidth results for the lowest \textit{p}
band near the BZ center calculated with the inclusion of the SO coupling term (see text for more details).}
\label{gamma_general}
\end{figure*}

\section{DAMPING RATES}

In Fig. \ref{gamma_general} we show the calculated damping rate
$\Gamma_{n\textbf{k}}=\tau^{-1}_{n\textbf{k}}$ of hot electrons and holes for all the values of band index $n$
and wave vectors \textbf{k} used in the calculations, both using the
scalar-relativistic energy spectrum (black dots) and including the SO coupling in the band structure (red dots). In addition, the blue dots represent the results obtained using the Lindhard dielectric function \cite{pino66} on calculating the screened interaction through Eq. (\ref{w_gg}) together
with the {\it ab initio} scalar-relativistic wave functions and one-particle energies in Eqs. (\ref{tau_init}) and (\ref{sigma_rr}). The
orange line is the result of a FEG model, i.e. when the band structure is described by a band with parabolic
dispersion, the wave functions are represented by plane-waves, and the Lindhard dielectric function is used for
the screening. In the inset of Fig. \ref{gamma_general}, the green curve represents the Quinn-Ferrel result
\cite{qufepr58} for the damping rate for very low energy quasiparticles:
\begin{equation}
\label{gamma_QF}
\tau_{QF}^{-1} = \frac{r_s^{5/2}(\varepsilon-E_F)^2}{263}eV^{-2}fs^{-1},
\end{equation}
where $r_s$ is the valence charge density parameter of the system. In the following we shall refer to these five kinds of lifetime calculations as SC, SO, LDF-SC, FEG, and QF ones, respectively.

There are three main features in the distribution of the damping rates in Fig. \ref{gamma_general}. First, for -8
eV $<\varepsilon<$ -6 eV the damping rates corresponding to the $s$-like holes are presented. On the opposite
side of the studied energy range, the values of $\Gamma_{n\textbf{k}}$ for excited electrons in the $d$-like
bands appear for energies in the 5.5 eV $<\varepsilon<$ 8 eV interval. However the main feature of Fig.
\ref{gamma_general} is the parabolic-like distribution of damping rate for the  $p$-like holes and electrons in
bands crossing the Fermi level. The linewidth data for the $p$ states is separated from that for the $s$-like
holes by the energy gap of $\approx$2.5 eV presented in the Pb band structure.

It is interesting to compare the observed difference on the calculated damping rates for $p$ and $d$ electrons
with the same energies. As can be seen in Fig. \ref{gamma_general}, at the same energy in the 5.5 eV
$<\varepsilon< $ 8 eV interval quasiparticles in the $d$ bands present a damping rate roughly two times smaller
in comparison with that for the $p$ bands. This is a consequence of the different coupling matrices [see Eq.
(10)] for \textit{p} and \textit{d} states thereby signalling about strong localization effect in lifetimes for $d$-states in Pb.

\begin{table}[t]
\caption{Linewidth for the lowest $p$ states at $\textbf{k}$'s in vicinity of the $\Gamma$ point along the
$\Gamma X$ direction, calculated at the scalar-relativistic level ($\Gamma^{\rm sc}$) and including the SO
splitting ($\Gamma^{\rm so}$) in the band structure. All values are in meV. The $\textbf{k}$ vectors are labeled
with the $\kappa$ index, corresponding to $\textbf{k}=\frac{2\pi}{24a_c}(\kappa,0,0)$, where $a_c$ is the lattice parameter.}
\begin{ruledtabular} \label{linewidth_G}
\begin{tabular}{crrrrrrr}
$\kappa$ & $0$ & $1$ & $2$ & $3$ & $4$ & $5$ & $6$ \\
\hline
$\Gamma^{\rm sc}$ & 1135 & 1100 & 1010 & 895 & 776 & 663 & 559 \\
$\Gamma^{\rm so}$ & 440 & 430 & 410
& 381 & 345 & 307 & 268 \\
\end{tabular}
\end{ruledtabular}
\end{table}

For quasiparticles in the $p$-like states, the role of final states in its decay [see Eq. (10)] can be seen in
Fig. \ref{gamma_general} from comparison of the distribution of the LDF-SC results (blue dots) with the FEG line.
As in both these cases the Lindhard dielectric function was employed, the difference arises only in the wave
functions entering the coupling matrix expression (\ref{b}). For the $p$ electrons and holes in the energy range 4
eV$\leqslant\varepsilon\leqslant$ 5 eV, both sets of data look very similar. Nevertheless, a more careful
analysis (presented below) shows some differences which are reflected in the calculated inelastic lifetimes.

The strongest SO effect in lifetime is observed for $p$ states in vicinity of the BZ center. In Fig.
\ref{gamma_general}, the arrow points to the $\Gamma_{n=2,\textbf{k}}$ results for $\textbf{k}$'s vectors close to
the $\Gamma$ point calculated with the inclusion of the SO splitting in the band structure. For completeness, in
Table \ref{linewidth_G} the linewidth value for the lowest $p$ band states at some $\textbf{k}$'s along the
$\Gamma X$ direction is presented.
These data demonstrate how the giant SO splitting of the $p$ bands at the $\Gamma$ point reduces the linewidth of
the states in this band by as much as $\approx60\%$. This is mostly due to the reduction of phase space for final states.

\begin{figure}[b]
\includegraphics[width=0.48\textwidth]{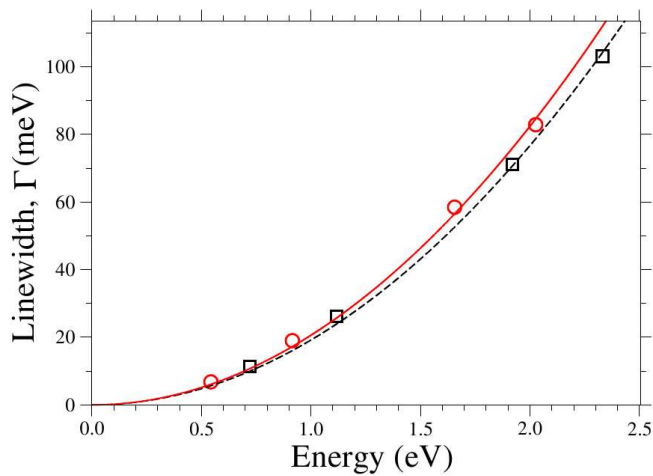}
\caption{(Color online) Calculated {\it ab initio} damping rates for excited electrons in states near the Fermi
level with wave vectors along the $\Gamma L$ direction excluding (squares, dashed line) and including (circles, solid line) SO interaction. Lines are fits by a quadratic function, $\Gamma=\alpha\cdotp\varepsilon^2$.}
\label{gamma_GL}
\end{figure}

Very recently lifetime measurements have been performed for the thin Pb films with (111) orientation grown on
semiconductors.\cite{hobrprb09,kirenp10} Such orientation corresponds to the quantization of bulk electronic
states along the $\Gamma L$ symmetry direction. Figure \ref{gamma_GL} presents the calculated damping rates for
the excited electrons in $p$ states along this direction. In difference with the situation for the electronic
states around the $\Gamma$ point one can see that the SO interaction produces small effect on lifetime for these
states. On the other hand, the presented calculated results are in good agreement with the experimental data for
thin films\cite{hobrprb09,kirenp10} signalling that low energy quasiparticles dynamics in such systems can be
well described by that in bulk Pb. Note also how the lifetime data for states along the $\Gamma L$ direction in
Fig. \ref{gamma_GL} both in the SC and SO calculations are well fitted by a quadratic function, $\Gamma=\alpha(E-E_F)^2$, over rather
extended energy range.

\begin{figure}[b]
\includegraphics[width=0.48\textwidth]{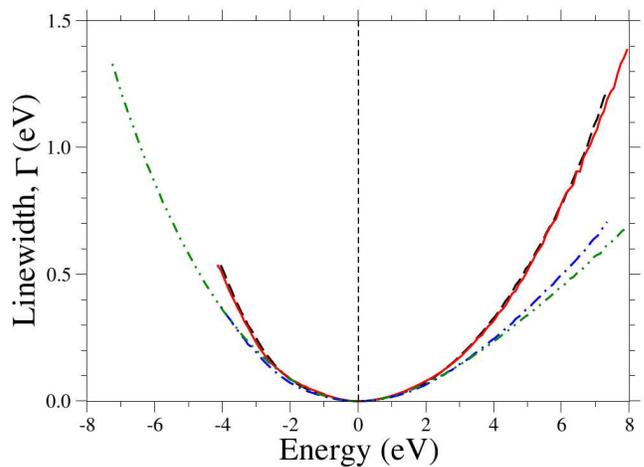}
\caption{(Color online) Energy dependence of averaged damping rates, $\Gamma(\varepsilon)$, of quasiparticles in
the \textit{p}-like bands obtained in the scalar-relativistic calculation (dashed line), the one including
the SO interaction (solid line) and the calculation which includes the {\it ab initio} eigenstates and the
Lindhard dielectric function (dashed-dotted line). The dashed-dotted-dotted line shows the same as
the orange solid line in Fig. \ref{gamma_general}.}
\label{gamma_averaged}
\end{figure}

\subsection{Dependence on the quasiparticle energy}

Although the damping rate $\Gamma_{n\textbf{k}}$ of a quasiparticle in a state ($n,{\bf k}$) depends on its band
index $n$ and wave vector $\textbf{k}$, one can define $\Gamma(\varepsilon)$ as an average of
$\Gamma_{n\textbf{k}}$ over states with all wave vectors and bands lying at the same energy in the BZ. In Fig.
\ref{gamma_averaged} the results for $\Gamma(\varepsilon)$ are presented for quasiparticles in the $p$-like
bands. The quasi-linear energy dependence of the averaged damping rate for holes with energies
$\varepsilon\leqslant$ -2.5 eV requires a separate analysis of the dependence of $\Gamma(\varepsilon)$ for hot
electrons and holes.

In Fig. \ref{gamma_averaged} different energy dependence of the averaged damping rate for holes is observed for
energies below and above $\varepsilon\approx$ -2.5 eV. At energies $\varepsilon$ above -2.5 eV, the averaged
linewidth presents a quasi-quadratic dependence on quasiparticle energy. For more detailed analysis of this
dependence, we fitted the $\Gamma(\varepsilon)$ data for -2.5 eV $\leqslant\varepsilon\leqslant 0$ by a
third-order polynomial:
\begin{equation}
\label{3fit}
\Gamma_{\rm fit}(\varepsilon) = a_1\cdotp\varepsilon+a_2\cdotp\varepsilon^2+a_3\cdotp\varepsilon^3.
\end{equation}
The coefficients $a_i$ obtained from this fitting are presented in Table \ref{fitting_coef} for all four kinds of
linewidth calculations. Note the highest value of the curvature in the SO case which is a direct consequence of
the avoiding band-crossing effect produced by inclusion of the SO coupling in the Hamiltonian seen in Fig.
\ref{Pb_band_structure}. In the -4.2 eV $\leqslant\varepsilon\leqslant$ -2.5 eV energy range the coefficients
$a_1\approx a_2\gg a_3$ for the SC, SO, LDF-SC calculated curves presenting strong contribution of a linear term
with exception of the FEG results which are well described by a quadratic function for any energy. Hence the
quasi-linear behavior of the averaged damping rates in all the calculations performed with the use of the {\it ab
initio} eigenstates reflects the non-free-electron like behavior of holes in the lowest $p$ band at energies
below -2.5 eV.

\begin{table*}
\caption{Fitting coefficients of the $\Gamma(\varepsilon)$ data using a third-order polynomial (\ref{3fit}) for
holes in the energy interval above -2.5 eV and \textit{p} electrons. Meaning of abbreviations "SC", "SO",
"LDF-SC", and "FEG" is explained in the text. $a_2$ in eV$^{-1}$ and $a_3$ in eV$^{-2}$.}
\begin{tabular}{m{.6cm}m{1.9cm}m{1.9cm}m{1.9cm}m{1.9cm}m{0.6cm}m{1.9cm}m{1.9cm}m{1.9cm}m{1.9cm}}
\hline\hline\noalign{\smallskip}
&\multicolumn{4}{c}{holes}&\multicolumn{1}{c}{}&\multicolumn{4}{c}{electrons}\\
\cline{2-5}\cline{7-10}\noalign{\smallskip}
&\multicolumn{1}{c}{SC}&\multicolumn{1}{c}{SO}&\multicolumn{1}{c}{LDF-SC}&\multicolumn{1}{c}{FEG}&\
\ &\multicolumn{1}{c}{SC}&\multicolumn{1}{c}{SO}&\multicolumn{1}{c}{LDF-SC}&\multicolumn{1}{c}{FEG}\\
\hline\noalign{\smallskip} $a_1$ & \multicolumn{1}{r}{$-0.0030$} & \multicolumn{1}{r}{$-0.0000$} &
\multicolumn{1}{r}{$-0.0020$} & \multicolumn{1}{r}{$0.0000$} & \ \ & \multicolumn{1}{r}{$0.0007$} &
\multicolumn{1}{r}{$-0.0020$} & \multicolumn{1}{r}{$0.0002$}
& \multicolumn{1}{r}{$0.0060$} \\
$a_2$ & \multicolumn{1}{r}{$0.0195$} & \multicolumn{1}{r}{$0.0313$} & \multicolumn{1}{r}{$0.0150$} &
\multicolumn{1}{r}{$0.0216$} & \ \ & \multicolumn{1}{r}{$0.0175$} & \multicolumn{1}{r}{$0.0190$}
& \multicolumn{1}{r}{$0.0180$} & \multicolumn{1}{r}{$0.0161$} \\
$a_3$ & \multicolumn{1}{r}{$0.0000$} & \multicolumn{1}{r}{$0.0050$} & \multicolumn{1}{r}{$-0.0009$} &
\multicolumn{1}{r}{$-0.0002$} & \ \ & \multicolumn{1}{r}{$0.0007$} & \multicolumn{1}{r}{$0.0004$} &
\multicolumn{1}{r}{$-0.0007$}
& \multicolumn{1}{r}{$-0.0008$} \\
\hline\hline
\end{tabular}\label{fitting_coef}
\end{table*}

In the case of electrons, all curves in Fig. \ref{gamma_averaged} present an apparent quadratic dependence on the
quasiparticle energy. Nevertheless, in this case we performed also the fitting procedure with the use of
expression (\ref{3fit}). Table \ref{fitting_coef} presents the obtained corresponding coefficients $a_i$ as well.
At first sight, all four studied curves show the expected quadratic energy dependence with the FEG results
presenting the greatest deviation. At the same time, from the data of Table \ref{fitting_coef} it is clear that
the SO curve has the strongest curvature (bigger quadratic coefficient).

An interesting point comes from the comparison of two $\Gamma(\varepsilon)$ curves calculated using the Lindhard
dielectric screening (LDF-SC and FEG cases), with the two curves calculated using the \textit{ab initio}
screening (SC and SO cases). The former ones deviate considerably for energies above $\sim 3$ eV from the two
latter. Nevertheless, the curvature (i.e., $a_2$ coefficients) is similar for all four curves (see Table
\ref{fitting_coef}). It is the negative sign of the $a_3$ coefficient in the case of the averaged damping rates
calculated with the Lindhard screening giving origin of the observed deviation in the high-energy side. Hence,
though $a_2\gg a_3$, for sufficiently high electron energies ($\varepsilon\geqslant$ $\sim3$ eV) the cubic term
in the dependence of $\Gamma(\varepsilon)$ on the energy can start to play significant role, the sign of $a_3$
being related to the screening.

\begin{table}
\caption{Effective charge density parameters obtained from Eq. (\ref{r_s}) on base of four sets of data as
explained in the text. $\delta r_s$ stands for the deviation from a conventional value for Pb $r_s=2.298$ a.u.}
\begin{ruledtabular} \label{r_ss}
\begin{tabular}{rrrrr}
 & SC & SO & LDF-SC & FEG \\
\hline\\
 $r_s^{\rm eff}$ & 2.37  &  2.49  &  2.18  &
 2.33  \\
 $\delta r_s^{\rm eff}(\%)$ &  +3.0  &  +8.3  &  -5.2  &
 +1.3  \\
\end{tabular}
\end{ruledtabular}
\end{table}

From the results for $\Gamma(\varepsilon)$ an effective charge density parameter $r_s^{\rm eff}$ can be derived
with the use of the Quinn-Ferrel expression (\ref{gamma_QF}). Fitting again the averaged linewidth curves with
Eq. (\ref{3fit}), new $a_i$ coefficients are calculated. Because of the approximations that lead to Eq.
(\ref{gamma_QF}) (see Refs. \onlinecite{ecpicp00} and \onlinecite{ecpicp01}), $\tau^{-1}_{QF}$ is only valid for very low energy quasiparticles. Thus, the new fitting is carried in the energy range -1 eV $\leqslant\varepsilon\leqslant$
1 eV, averaging the effect of possible different curvatures for electrons and holes. Finally, the $r_s^{\rm eff}$
parameters are found using the following expression:
\begin{equation}
\label{r_s}
r_s^{\rm eff} = (399.7\times a_2)^{2/5}.
\end{equation}
In Table \ref{r_ss} the calculated $r_s^{\rm eff}$ and their deviation from the conventional $r_s=2.298$ a.u. for
Pb are presented. Taking into account that fitting details, i.e., the choice of the polynomial or of the energy
range applied in the fitting procedure, may influence the exact resulting values of $r_s^{\rm eff}$, a good
agreement with $r_s$ is found. The value of the effective density parameter in the FEG case is the closest one
whereas the higher value of $r_s^{\rm eff}$ for the SO case in comparison with the purely scalar-relativistic one
means that the effective screening at such low quasiparticle energies is stronger in the former case.

\begin{table}
\caption{Lifetime of excited electrons (holes) at four different values of $|\varepsilon|$. All values are in
fs.}
\begin{ruledtabular} \label{lifetime}
\begin{tabular}{crrrrr}
$|\varepsilon|$  & SC &  SO & LDF-SC & FEG & QF \\
\hline\\
 0.5 eV& 130(114)& 114(87) & 162(145) & 128(130) &  133 \\
 1.0 eV&   34(29)&  31(23) &   41(36) &   34(31) &   33 \\
 2.0 eV&     9(8)&    8(7) &    10(9) &     9(7) &    8 \\
 3.0 eV&     4(3)&    4(3) &     4(3) &     5(3) &    4 \\
\end{tabular}
\end{ruledtabular}
\end{table}

In Table \ref{lifetime} we present the calculated values for averaged inelastic lifetimes for excited electrons
and holes in Pb at $|\varepsilon|=$ 0.5, 1.0, 2.0 and 3.0 eV obtained in all the calculations. For excited
electrons, the QF, FEG, and SC calculations give similar results. Noting that the calculated lifetimes using the
self-consistent eigenstates and the Lindhard screening differ from that three ones, we can conclude that, even if
lead is a free-electron-like metal,  lifetime of quasiparticles in bulk Pb is the result of a balance between
screening and localization. Interestingly, the lifetimes in this energy interval calculated including the SO
splitting in the band structure are the lowest ones for each quasiparticle energy, both for electrons and holes.

\begin{figure}[t]
\includegraphics[width=0.48\textwidth]{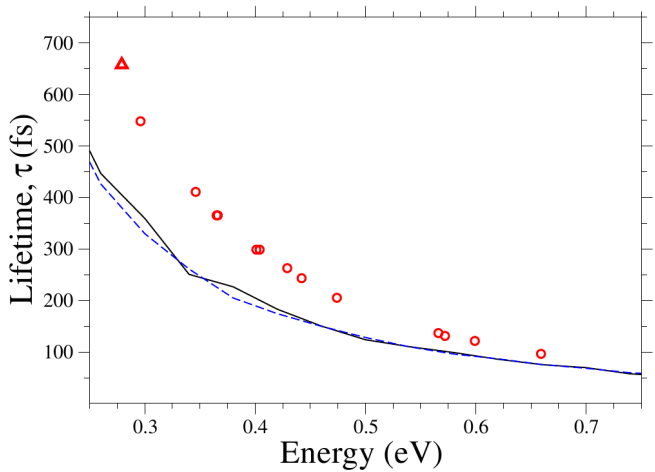}
\caption{(Color online) Averaged inelastic lifetimes of electrons (solid line) and the lifetimes calculated for
states at \textbf{k}'s in vicinity of the $W$ point (circles) evaluated in the scalar-relativistic calculation.
Triangle highlights $\tau$ at the W point. Dashed line: FEG results.}
\label{lifetime_W}
\end{figure}

As seen in Fig. \ref{Pb_band_structure}, at the $W$ point only one unoccupied $p$ band approaches but does not
cross $E_F$, presenting a local minimum. This proximity to the Fermi level together with the absence of available
unoccupied states for decay with small momentum transfers leads to notably longer inelastic lifetimes for the
states around the $W$ point. This is a strong band structure effect. In Fig. \ref{lifetime_W} we compare the
averaged inelastic lifetimes, $\tau^{\rm av}$, of electrons with very low energies (solid line) with the values
calculated for states close to $W$ (shown by symbols) obtained in the scalar-relativistic calculation (note that
for this analysis the SO interaction has negligible effect). For comparison, the FEG results are shown by dashed
line. As can be seen, upon approaching the $W$ point (reducing energy) $\tau$ increases faster in comparison with
the averaged inelastic lifetime at those energies, being by $\approx 50\%$ higher than $\tau^{\rm av}$ at the
same energy in other parts of the BZ. Hence, around the $W$ point, $\tau$ behaves in a non-free-electron-like
manner.

\begin{figure}[t]
\includegraphics[width=0.48\textwidth]{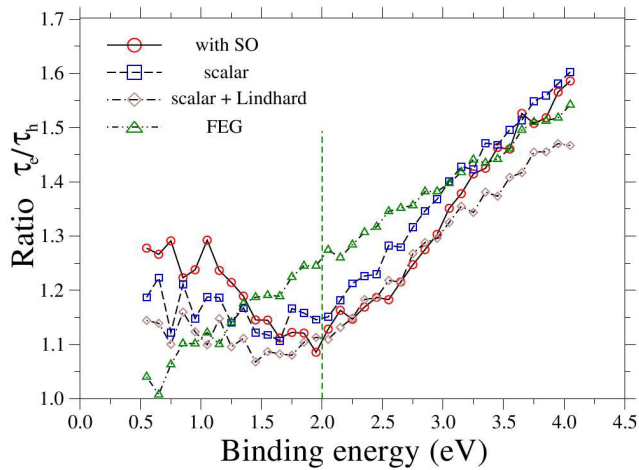}
\caption{(Color online) Ratio $\tau_{e}/\tau_{h}$ as a function of the quasiparticle energy for four levels of
calculations. The dashed line separates two energy regions with different physical behavior (see text). Lines are
guides to the eye.}
\label{lifetime_ratio}
\end{figure}

\subsection{Ratio of the lifetime of electrons and holes}

In Fig. \ref{lifetime_ratio}, the ratio of lifetimes for electrons, $\tau_e$, and holes, $\tau_h$, at the same
absolute value of the quasiparticle energy are shown for four different calculations. The curves show two energy
ranges with different behavior. At energies $|\varepsilon|\geqslant$2 eV all the four curves present a
quasi-linear behavior. However, for quasiparticle energies below 2 eV, only the FEG curve remains quasi-linear.
The other three calculations present quite different dependence of the $\tau_{e}/\tau_{h}$ ratio at those
energies. The quasi-linear behavior of $\tau_{e}/\tau_{h}$ is found in the homogeneous electron gas calculations
(see ). Hence in bulk Pb the band structure effects are
important in the electron-electron inelastic scattering processes for quasiparticles with energies less than 2
eV. As the LDF-SC curve deduced from the lifetime results obtained with the use the Lindhard screening and the
{\it ab initio} eigenstates presents also the band structure effects, these effects are the consequence of using
the true eigenstates in the evaluation of the coupling matrices [see Eq. (10)], and not of the {\it ab initio}
screening used. Note also that $\tau_{e}>\tau_{h}, \forall|\varepsilon|$, for all four levels of calculations.

\begin{figure}[b]
\includegraphics[width=0.49\textwidth]{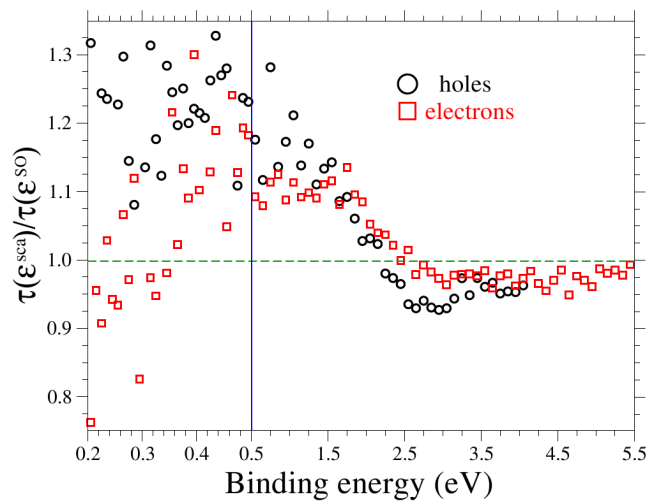}
\caption{(Color online) Ratio $\tau(\varepsilon^{scalar})/\tau_(\varepsilon^{SO})$ as a function of the
quasiparticle energy for \textit{p}-like states, both for holes (circles) and electrons (squares). The
vertical line separates the quasiparticle energy ranges with different scales.}
\label{lifetime_ratio_so}
\end{figure}


\subsection{Effect of SO interaction on lifetime}

Modifications in the Pb band structure upon inclusion of the SO interaction increase lifetime of electrons in the
\textit{d} bands and reduce that of holes in the \textit{s} ones, as expected from Fig. \ref{gamma_general}. In
all the cases the variation in $\tau$ upon inclusion of the SO term does not exceed $10\%$ in comparison with
values obtained in the scalar-relativistic calculations. In Fig. \ref{lifetime_ratio_so} the ratio
$\tau(\varepsilon^{\rm SC})/\tau_(\varepsilon^{\rm SO})$ as a function of energy is plotted for the \textit{p}
holes and electrons.
For holes with $\varepsilon\leqslant-2$ eV and electrons with 2.5 eV $\leqslant\varepsilon\leqslant$ 5.5 eV
inclusion of the SO splitting increases the quasiparticle lifetimes, in both cases less than $10\%$. On the other
hand, for holes with binding energies less than 2 eV, spin-orbit induced splitting reduces lifetimes as much as
by $30\%$. For electrons with $\varepsilon\leqslant$ 2.5 eV, as can be seen in Fig. \ref{lifetime_ratio_so}, the
SO interaction lowers $\tau$ by $\approx10\%$. One can expect that this tendency will continue for electronic
states around the Fermi surface. However, the {\bf k} mesh used in the samplings over the BZ in present lifetime
calculations does not allow to obtain reliable lifetime values for the states with binding energies less than 0.5
eV.

\section{CONCLUSIONS}
We have presented GW-RPA calculations of the inelastic damping rates of low energy quasiparticles in bulk Pb
and studied in detail the band structure effects as well as the consequences of the inclusion of the spin-orbit
interaction. A strong localization effect is found for electrons with energies 5.5 eV
$\leqslant\varepsilon\leqslant$ 8 eV, where lifetime for $d$ states is almost 2 times longer than that for $p$
states.

The states in the lowest valence $p$ energy band at the center of the BZ reduce their damping rates by roughly
$\approx60\%$ upon inclusion of the spin-orbital interaction in comparison with the scalar-relativistic
calculation.
For the $p$ electrons the damping rates $\Gamma(\varepsilon)$ averaged over the BZ present a
quadratic dependence on energy $\varepsilon$, whereas for $p$ holes the quasi-linear dependence of
$\Gamma(\varepsilon)$ at $\varepsilon$ below -2.5 eV reflects a non-free-electron-like nature of the electronic
states at the bottom of the lowest $p$ band.
The calculated lifetimes ratio $\tau_e(\varepsilon)/\tau_h(\varepsilon)$ reveals that band structure effects are
important in electron-electron inelastic scattering processes for quasiparticles with binding energies smaller than
2 eV.
Present lifetime calculations for bulk Pb are in good agreement with inelastic damping rates of quantum
well states of metallic thin films measured very recently presenting thereby evidence that quasiparticle dynamics
can be considered as being bulk-like even in very thin Pb films.

\section*{ACKNOWLEDGEMENTS}

We acknowledge financial support from the Spanish MICINN (No. FIS2007-66711-C02-01), the Departamento de
Educaci\'on del Gobierno Vasco, and the University of the Basque Country (No. GIC07-IT-366-07).


\begin{thebibliography}{99}

\bibitem{qufepr58} J. J. Quinn and R. A. Ferrell, Phys. Rev. {\bf 112}, 812 (1958).

\bibitem{qupr62} J. J. Quinn, Phys. Rev. {\bf 126}, 1453 (1962).

\bibitem{no62} P. Nozi\`eres, \textit{Theory of Interacting Fermi Liquids} (Benjamin, New York, 1962).

\bibitem{capiprl99} I. Campillo, J. M. Pitarke, A. Rubio, E. Zarate, and P. M.
Echenique, Phys. Rev. Lett. {\bf 83}, 2230 (1999).

\bibitem{sckeprb99} W.-D. Sch{\"o}ne, R. Keyling, M. Bandi\'c, and W. Ekardt, Phys. Rev. B {\bf 60}, 8616 (1999).

\bibitem{kescprb00}  R. Keyling, W.-D. Sch{\"o}ne, and W. Ekardt, Phys. Rev. B {\bf 61}, 1670 (2000).

\bibitem{caruprl00} I. Campillo, A. Rubio, J. M. Pitarke, A. Goldmann, and P. M. Echenique, Phys. Rev. Lett. {\bf 85}, 3241 (2000).

\bibitem{capiprb00} I. Campillo, J. M. Pitarke, A. Rubio, and P. M. Echenique, Phys. Rev. B {\bf 62}, 1500 (2000).

\bibitem{gebeprb01} A. Gerlach, K. Berge, A. Goldmann, I. Campillo, A. Rubio, J. M. Pitarke, and P. M. Echenique, Phys. Rev. B {\bf 64}, 085423 (2001).

\bibitem{ladeprb03} F. Ladst{\"a}dter, P. F. de Pablos, U. Hohenester, P. Puschnig, C. Amdrosch-Draxl, P. L. de Andres, F.J.Garc\'ia Vidal, and F. Flores, Phys. Rev. B {\bf 68}, 085107 (2003).

\bibitem{pizhchemphys04} J. M. Pitarke, V. P. Zhukov, R. Keyling, E. V. Chulkov, and P. M. Echenique, ChemPhysChem {\bf 5}, 1284 (2004). 

\bibitem{lahoprb04} F. Ladst{\"a}dter, U. Hohenester, P. Puschnig, and C. Amdrosch-Draxl,
Phys. Rev. B {\bf 70}, 235125 (2004).

\bibitem{zhchpu09} V. P. Zhukov and E. V. Chulkov, Phys. Usp. {\bf 52}, 105 (2009) [Usp. Fiz. Nauk {\bf 179} (2), 113
(2009)].

\bibitem{nezhpss07} I. A. Nechaev, V. P. Zhukov, and E. V. Chulkov, Phys. Solid State {\bf 49}, 1811 (2007)
[Fiz. Tverd. Tela (S.-Petersburg) {\bf 49}, 1729 (2007)].

\bibitem{casiprb00}  I. Campillo, V. M. Silkin, J. M. Pitarke, E. V. Chulkov, A. Rubio, and P. M. Echenique,
Phys. Rev. B {\bf 61}, 13484 (2000).

\bibitem{neskprb08} I. A. Nechaev, I. Yu. Sklyadneva, V. M. Silkin, P. M. Echenique, and E. V. Chulkov, Phys.
Rev.  B {\bf 78} 085113 (2008).

\bibitem{sichprb03} V. M. Silkin, E. V. Chulkov,  and P. M. Echenique, Phys. Rev. B {\bf 68}, 205106 (2003).

\bibitem{ekscapa00} W. Ekardt, W.-D. Sch{\"o}ne, and R. Keyling,  Appl. Phys. A: Mater. Sci. Process. {\bf 71}, 529 (2000).

\bibitem{scijmp03} W.-D. Sch{\"o}ne, Int. J. Mod. Phys. {\bf 17}, 5655 (2003).

\bibitem{gupiprb04} I. G. Gurtubay, J. M. Pitarke, and P. M. Echenique, Phys. Rev. B {\bf 69},
245106 (2004).

\bibitem{zharprb01}  V. P. Zhukov, F. Aryasetiawan, E. V. Chulkov, I. G. de Gurtubay, and P. M. Echenique,
Phys. Rev. B {\bf 64}, 195122 (2001).

\bibitem{bascprb02} M. R. Bacelar, W.-D. Sch\"one, R. Keyling, and W. Ekardt, Phys. Rev. B {\bf 66},
153101 (2002).

\bibitem{scpss07} W.-D. Sch\"one, Prog. Surf. Sci. {\bf 82}, 161 (2007).

\bibitem{zhchprl04}  V. P. Zhukov, E. V. Chulkov, and P. M. Echenique, Phys. Rev. Lett. {\bf 93}, 096401 (2004).

\bibitem{appphya} V. M. Silkin, A. Balassis, A. Leonardo, E. V. Chulkov, and P. M. Echenique, Appl. Phys. A: Mater. Sci. Process. \textbf{92}, 453 (2008).

\bibitem{nechpss09} I. A. Nechaev and E. V. Chulkov, Phys. Solid State {\bf 51}, 754 (2009)
[Fiz. Tverd. Tela (S.-Petersburg) {\bf 51}, 713 (2009)].

\bibitem{nechepjb10} I. A. Nechaev and E. V. Chulkov, Eur. Phys. J. B (2010) (in press).

\bibitem{helussp69}  L. Hedin and S. Lundqvist, Solid State Phys. {\bf 23}, 1 (1969).

\bibitem{fewa71}  A. L. Fetter  and J. D. Walecka, \textit{Quantum Theory of Many-Particle Systems}
(McGraw-Hill, Boston, 1971).

\bibitem{ma90}  G. D. Mahan, \textit{Many-Particle Physics} (Plenum Press, New York, 1990).

\bibitem{zhanprb04}  V. P. Zhukov, O. Andreyev, D. Hofmann, M. Bauer, M. Aeschlimann, E. V. Chulkov,
and P. M. Echenique, Phys. Rev. B {\bf 70}, 233106 (2004).

\bibitem{zhchprb05}  V. P. Zhukov, E. V. Chulkov, and P. M. Echenique, Phys. Rev. B {\bf 72}, 155109 (2005).

\bibitem{zharprb02}  V. P. Zhukov, F. Aryasetiawan, E. V. Chulkov, and P. M. Echenique, Phys. Rev. B {\bf 65},
115116 (2002).

\bibitem{madeprb02} A. Marini, R. Del Sole, A. Rubio, and G. Onida,
Phys. Rev. B {\bf 66}, 161104 (2002).

\bibitem{yimaprb10} Z. Yi, Y. Ma, M. Rohlfing, V. M. Silkin, and  E. V. Chulkov,
Phys. Rev. B {\bf 81}, 125125 (2010).




\bibitem{guzhs04} Y. Guo, Y.-F. Zhang, X.-Y. Bao, T.-Z. Han, Z. Tang, L.-X. Zhang, W.-G. Zhu, E. G. Wang, Q. Niu,
Z. Q. Qiu, J.-F. Jia, Z.-X. Zhao, and Q.-K. Xue, Science {\bf 306}, 1915 (2004).

\bibitem{eoqiprl06} D. Eom, S. Qin, M.-Y. Chou, and C. K. Shih, Phys. Rev. Lett. {\bf 96}, 027005 (2006).

\bibitem{qikis09} S. Qin, J. Kim, Q. Niu, and C.-K. Shih, Science {\bf 324}, 1314 (2009).

\bibitem{brhoprl09} C. Brun, I-P. Hong, F. Patthey, I. Y. Sklyadneva, R. Heid, P. M. Echenique, K. P. Bohnen, E.
V. Chulkov, and W.-D. Schneider, Phys. Rev. Lett. {\bf 102}, 207002 (2009).

\bibitem{zhchnp10} T. Zhang, P. Cheng, W.-J. Li, Y.-J. Sun, G. Wang, X.-G. Zhu, K. He, L. Wang X. Ma, X. Chen, Y. Wang, Y. Liu, H.-Q. Lin, J.-F. Jia, and Q.-K. Xue, Nat. Phys. {\bf 6}, 104 (2010).

\bibitem{hobrprb09}  I-P. Hong, C. Brun, F. Patthey, I. Yu. Sklyadneva, X. Zubizarreta, R. Heid,
V. M. Silkin, P. M. Echenique, K. P. Bohnen, E. V. Chulkov, and W.-D. Schneider, Phys. Rev. B {\bf 80}, 081409(R)
(2009).
\bibitem{kirenp10} P. S. Kirchmann, L. Rettig, X. Zubizarreta, V. M. Silkin, E. V. Chulkov, and U. Bovensiepen,
Nat. Phys. {\bf 6} (2010) (in press).

\bibitem{hokopr64}  P. Hohenberg and W. Kohn, Phys. Rev. {\bf 136}, B864 (1964).

\bibitem{koshpr65}  W. Kohn and L. Sham, Phys. Rev. {\bf 140}, A1133 (1965).

\bibitem{rugrprl84}  E. Runge and E. K. U. Gross, Phys. Rev. Lett. {\bf 52}, 997 (1984).

\bibitem{pegoprl96}  M. Petersilka, U. J. Gossmann, and E. K. U. Gross, Phys. Rev. Lett. {\bf 76}, 1212 (1996).

\bibitem{adpr62} S. L. Adler, Phys. Rev. {\bf 126}, 413 (1962).

\bibitem{trmaprb91} N. Troullier and J. L. Martins, Phys. Rev. B {\bf 43}, 1993 (1991).

\bibitem{cealprl80}  D. M. Ceperley and B. J. Alder, Phys. Rev. Lett. {\bf 45}, 566 (1980).

\bibitem{pezuprb81}  J. P. Perdew and A. Zunger, Phys. Rev. B {\bf 23}, 5048 (1981).

\bibitem{arguprb94} F. Aryasetiawan and O. Gunnarsson, Phys. Rev. B {\bf 49}, 16214 (1994).

\bibitem{sichprb09} V. M. Silkin, I. P. Chernov, Yu. M. Koroteev, and E. V. Chulkov,
Phys. Rev. B {\bf 80}, 245114 (2009).

\bibitem{ti71}  M. Tinkam, \textit{Group Theory and Quantum Mechanics} (McGraw-Hill, New York, 1971).

\bibitem{vetoprb08}  M. J. Verstraete, M. Torrent, F. Jollet, G. Z\'erah, and X. Gonze, Phys. Rev. B {\bf 78}, 045119 (2008).

\bibitem{jepoprb90}  G. J\'ez\'equel and I. Pollini, Phys. Rev. B {\bf 41}, 1327 (1990).

\bibitem{Bi_SO}  X. Gonze, J.-P. Michenaud, and J.-P. Vigneron, Phys. Rev. B {\bf 41}, 11827 (1990).

\bibitem{pino66}  D. Pines  and P. Nozi\`eres, \textit{The Theory of Quantum Liquids} (Benjamin, New York, 1966).

\bibitem{ecpicp00} E. V. Chulkov, A. G. Borisov, J. P. Gauyacq, D. S\'anchez-Portal, V. M. Silkin, V. P. Zhukov, and P. M. Echenique, Chem. Rev. {\bf 106}, 4160 (2006), and references therein.

\bibitem{ecpicp01} P. M. Echenique, R. Berndt, E. V. Chulkov, Th. Fauster, A. Goldmann, and U. H{\"o}fer, Surf. Sci. Rep. {\bf 52}, 219 (2004), and references therein.







\end{thebibliography}
\end{document}